\documentclass[twocolumn]{aastex63}

\usepackage{times}
\usepackage{amsmath}
\usepackage{graphicx}
\usepackage{hyperref}
\usepackage{gensymb}
\usepackage{upgreek}
\usepackage{natbib}
\usepackage{subfigure}
\usepackage{multirow}
\usepackage{comment}
\usepackage{mathtools}
\usepackage{mathrsfs}
\usepackage{fontenc}
\usepackage{color}
\usepackage{url}
\usepackage{pifont}

\newcommand{\Ha}{H$\alpha$}
\newcommand{\Hb}{H$\beta$}
\newcommand{\NeII}{[Ne\,\textsc{ii}]}
\newcommand{\NeIII}{[Ne\,\textsc{iii}]}
\newcommand{\NeV}{[Ne\,\textsc{v}]}
\newcommand{\OII}{[O\,\textsc{ii}]}
\newcommand{\OIII}{[O\,\textsc{iii}]}

\submitjournal{ApJ}

\shorttitle{}
\shortauthors{Zhuang et al.}

\begin{document}

\title{Black Hole Accretion Correlates with Star Formation Rate and Star Formation Efficiency in Nearby Luminous Type 1 Active Galaxies}

\author[0000-0001-5105-2837]{Ming-Yang Zhuang}
\email{mingyangzhuang@pku.edu.cn}
\affil{Kavli Institute for Astronomy and Astrophysics, Peking University,
Beijing 100871, China}
\affil{Department of Astronomy, School of Physics, Peking University,
Beijing 100871, China}

\author[0000-0001-6947-5846]{Luis C. Ho}
\affil{Kavli Institute for Astronomy and Astrophysics, Peking University,
Beijing 100871, China}
\affil{Department of Astronomy, School of Physics, Peking University,
Beijing 100871, China}

\author[0000-0002-4569-9009]{Jinyi Shangguan}
\affil{Max-Planck-Institut f\"{u}r extraterrestrische Physik, 
Gie{\ss}enbachstr. 1, D-85748 Garching, Germany}

\begin{abstract}
We investigate the relationship between black hole accretion and star formation in a sample of 453 $z\approx0.3$ type~1 active galactic nuclei (AGNs).  We use available CO observations to demonstrate that the combination of nebular dust extinction and metallicity provides reliable estimates of the molecular gas mass even for the host galaxies of type~1 AGNs.  Consistent with other similar but significantly smaller samples, we reaffirm the notion that powerful AGNs have comparable gas content as nearby star-forming galaxies and that AGN feedback does not deplete the host of cold gas instantaneously. We demonstrate that while the strong correlation between star formation rate and black hole accretion rate is in part driven by the mutual dependence of these parameters on molecular gas mass, the star formation rate and black hole accretion rate are still weakly correlated after removing the dependence of star formation rate on molecular gas mass. This, together with a positive correlation between star formation efficiency and black hole accretion rate, may be interpreted as evidence for positive AGN feedback. 
\end{abstract}


\section{Introduction} \label{sec1}

Active galactic nuclei (AGNs) play a significant role in galaxy evolution by changing the ionization structure and injecting energy and momentum into the interstellar medium. The fast outflows produced by AGNs are expected to heat and sweep out most of the gas in their host galaxies, inhibiting star formation and preventing the galaxies from overgrowing, in what is often dubbed ``negative'' feedback \citep[see][for a review]{2012ARA&A..50..455Fabian}. On the other hand, the compression of gas in the interstellar medium \citep{2005ApJ...635L.121King, 2013ApJ...772..112Silk} and direct formation of stars inside outflows \citep{2017Natur.544..202Maiolino+, 2018NewA...61...95Wang&Loeb, 2019MNRAS.485.3409Gallagher+} could enhance the star formation rate (SFR) through ``positive'' feedback. However, negative and positive feedback do not necessarily act against each other.  They sometimes occur simultaneously in the same galaxy \citep{2013ApJ...772..112Silk, 2015ApJ...799...82Cresci+}. The complex interplay between AGNs and their large-scale environment lies at the heart of the coevolution of supermassive black holes (BHs) and their host galaxies \citep{Richstone1998, 2013ARA&A..51..511Kormendy&Ho, 2014ARA&A..52..589Heckman+}. 

Much effort has been invested in elucidating the link between the SFR of AGN host galaxies and the luminosity of the AGN, or, equivalently, the accretion rate of the central BH ($\dot{M}_{\rm BH}$). Some studies report that SFR strongly correlates with $\dot{M}_{\rm BH}$ \citep[e.g.,][]{2012ApJ...753L..30Mullaney+, 2013ApJ...773....3Chen+, 2016MNRAS.457.4179Harris+, 2017A&A...602A.123Lanzuisi+, Zhuang&Ho2020}, while others find a shallower correlation, or none at all \citep[e.g.,][]{2015ApJ...806..187Azadi+, 2015MNRAS.453..591Stanley+, 2017MNRAS.472.2221Stanley+, 2017MNRAS.466.3161Shimizu+}. The relation beween SFR and $\dot{M}_{\rm BH}$ may also depend on luminosity and redshift \citep[e.g.,][]{2010ApJ...712.1287Lutz+, 2012A&A...545A..45Rosario+, 2012A&A...540A.109Santini+}. Many factors may contribute to these contradictory results, including sample size, sample selection, and the methods used to calculate the SFR and to bin the data \citep{2017NatAs...1E.165Harrison}. 

Measuring accurate SFRs in AGN host galaxies presents a major observational challenge.  The emission from rapidly accreting BHs can easily dominate the observed spectral energy distribution and contaminate traditional SFR diagnostics normally employed in star-forming galaxies. Thermal dust emission from the obscuring torus and the narrow-line region can contribute significantly to the infrared (IR) continuum \citep[e.g.,][]{2006A&A...458..405Groves+, 2017MNRAS.466.3161Shimizu+, 2018ApJ...854..158Shangguan+, 2018ApJ...862..118Zhuang+}, which is otherwise widely used to measure the SFR of inactive galaxies \citep{Kennicutt1998}. While polycyclic aromatic hydrocarbons closely trace ultraviolet photons from young stars \citep{2016ApJ...818...60Shipley+, 2019ApJ...884..136Xie&Ho}, they can be destroyed by the more intense, harder radiation field of AGNs \citep{Voit1992, 2020NatAs...4..339Li}. Many attempts have been made to derive more reliable SFR diagnostics in AGNs, ranging from developing more sophisticated models of the IR emission \citep[e.g.,][]{2017ApJ...838L..20Honig&Kishimoto, 2017ApJ...841...76Lyu&Rieke, 2019MNRAS.484.3334Stalevski+}, improving the methods  for fitting the spectral energy distribution \citep[e.g.,][]{2015A&A...576A..10Ciesla+, 2020MNRAS.491..740Yang+}, and devising empirical calibrations based on certain diagnostic emission lines \citep[e.g.,][]{2005ApJ...629..680Ho, 2008ApJ...689...95Melendez+, 2010ApJ...725.2270Pereira-Santaella+, 2016MNRAS.462.1616Davies+, 2018ApJ...856...89Thomas+}.  Building upon \cite{HoKeto2007}, \cite{2019ApJ...873..103Zhuang+} used photoionization models that employ realistic AGN spectral energy distributions and physical properties of the narrow-line region to calibrate a new SFR estimator for AGNs anchored on the mid-IR fine-structure lines of \NeII~$12.81\,\micron$, \NeIII~$15.55\,\micron$, and \NeV~$14.32\, \micron$.  The same set of models was then extended by \cite{2019ApJ...882...89Zhuang&Ho} to the optical lines of \OII~$\lambda3727$ and \OIII~$\lambda5007$, updating the prior effort of \cite{Kim2006}. 

A positive relation between two variables does not necessarily signify an underlying causal connection if the correlation is artificially driven by the mutual dependence of the two parameters on a third. This may be a source of concern for the reported ${\rm SFR}-\dot{M}_{\rm BH}$ correlation, especially when it derives from AGN samples covering a wide range of redshift.  A number of studies have shown that the correlation is significantly reduced after accounting for the redshift dependence \citep[e.g.,][]{2012A&A...545A..45Rosario+, 2015MNRAS.453..591Stanley+, 2017A&A...602A.123Lanzuisi+, 2018MNRAS.478.4238Dai+}.  The separate dependence of SFR and $\dot{M}_{\rm BH}$ on stellar masses poses a similar ambiguity \citep{2017ApJ...842...72Yang+, 2019ApJ...872..168Suh+, 2020ApJ...888...78Stemo+}.  Here we focus on yet another factor---the impact of molecular gas mass ($M_{\rm H_2}$).  As the raw material that directly forms stars, $M_{\rm H_2}$ strongly correlates with SFR, both as integrated on global scales and as resolved on sub-galactic scales \citep[e.g.,][]{1998ApJ...498..541Kennicutt, 2008AJ....136.2846Bigiel+, 2010MNRAS.407.2091Genzel+}. The integrated molecular gas content of the host galaxies of nearby AGNs is also found to correlate with $\dot{M}_{\rm BH}$ \citep{2012ApJ...750...92Xia+, 2017MNRAS.470.1570Husemann+}, as might arise if the large-scale interstellar medium of the host couples with the fuel supply on circumnuclear scales.  In their analysis of 40 low-redshift ($z<0.3$) quasars with CO and far-IR observations, \citet{Shangguan+2020} show that while their sample exhibits a statistically significant correlation between IR luminosity and AGN luminosity, or, equivalently, a relation between SFR and $\dot{M}_{\rm BH}$, the relation vanishes once the mutual correlation of the two quantities with CO luminosity is removed.  One of the main objectives of the present study is to test the robustness of this result, which we achieve by substantially expanding the sample by an order of magnitude.

Cold molecular gas in galaxies is usually traced using CO \citep{YoungScoville1991}, and more recently [C~I] \citep[e.g.,][]{Valentino2018}, but observations of these lines are expensive and difficult to acquire for large, representative samples of AGNs.  Alternative methods of estimating the cold interstellar medium have been developed based dust emission in the thermal IR \citep{2007ApJ...663..866Draine+, 2014ApJ...783...84Scoville+} and dust attenuation derived from optical hydrogen Balmer lines \citep[e.g.,][]{2013MNRAS.432.2112Brinchmann+, 2019MNRAS.486L..91Concas&Popesso}. Recently, \citet{2019ApJ...884..177Yesuf&Ho} proposed an effective formalism to predict molecular gas mass from optical nebular dust extinction and metallicity, which enables efficient estimates of molecular gas mass for large spectroscopic surveys of galaxies. 

Taking advantage of the recent \OII~$\lambda 3727$ SFR estimator for AGNs developed by \citet{2019ApJ...882...89Zhuang&Ho}, \citet{Zhuang&Ho2020} assembled a large sample of low-redshift type~1 AGNs to systematically investigate the star formation properties of the host galaxies and their relation to their accreting BHs.  After properly mitigating the influence of redshift, \citet{Zhuang&Ho2020} show that their AGN sample still exhibits a strong correlation between SFR and $\dot{M}_{\rm BH}$. They find no obvious dependence on stellar mass. Here we extend the analysis a step further by considering the possible effect of molecular gas mass, which we obtain using the method of \citet{2019ApJ...884..177Yesuf&Ho}.  This paper assumes a cosmology with $H_0=70$ km~s$^{-1}$~Mpc$^{-1}$, $\Omega_m=0.3$, and $\Omega_{\Lambda}=0.7$.  We adopt the stellar initial mass function of \citet{2001MNRAS.322..231Kroupa}.

\begin{figure*}[t]
\centering
\includegraphics[width=\textwidth]{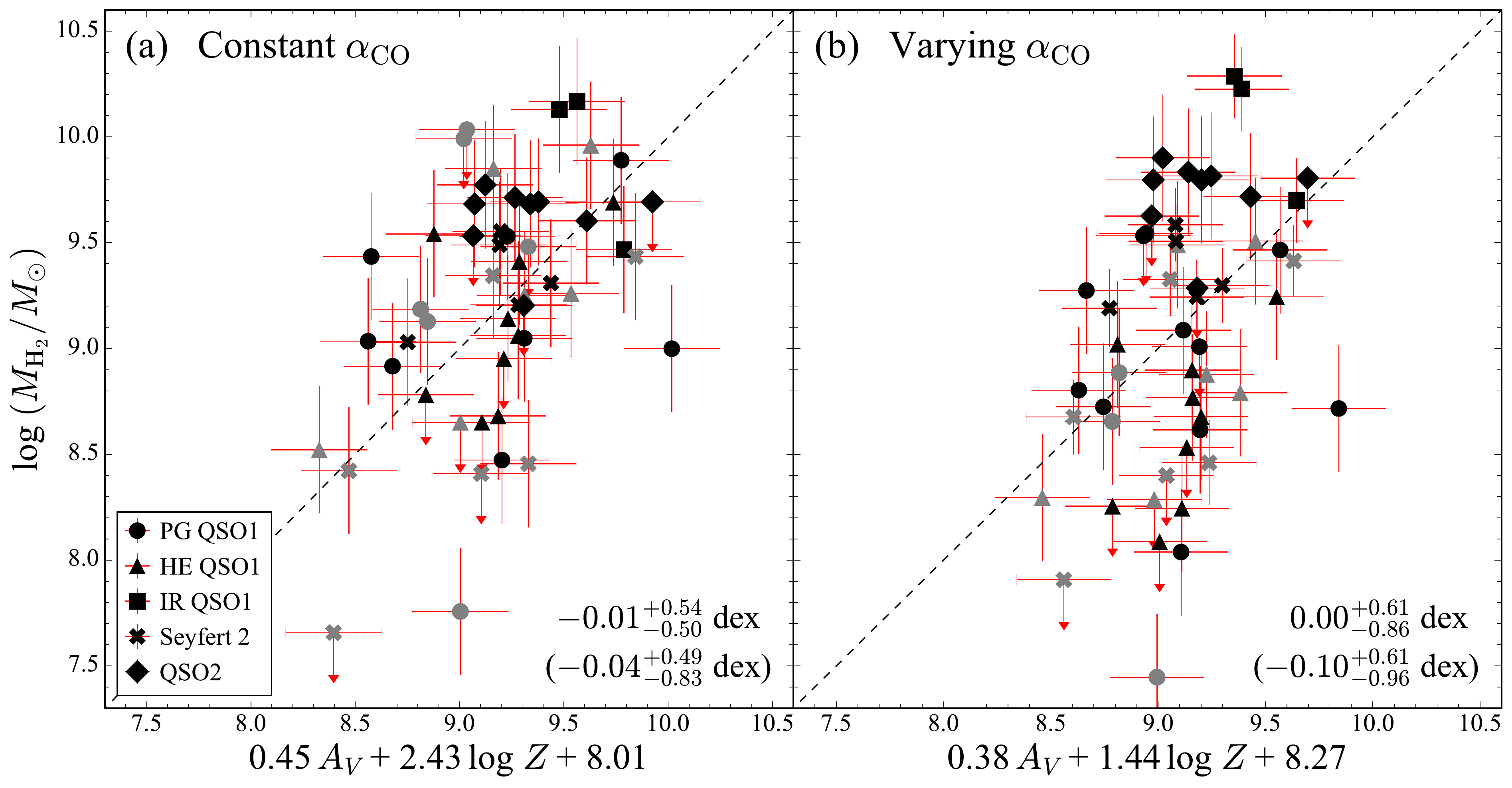}
\caption{Comparison of molecular gas masses estimated from CO(1--0) measurements with those estimated from dust extinction and metallicity \citep{2019ApJ...884..177Yesuf&Ho}, assuming (a) constant and (b) varying CO-to-H$_2$ conversion factor $\alpha_{\rm CO}$. We assume a constant $\alpha_{\rm CO}=3.1\ M_{\sun}$~(K km s$^{-1}$~pc$^2$)$^{-1}$ for the Palomar-Green (PG) quasars, Hamburg/ESO (HE) quasars, and Seyfert 2 galaxies; for the IR-luminous quasars and type~2 quasars, we adopt $\alpha_{\rm CO}=0.8\ M_{\sun}$~(K km s$^{-1}$~pc$^2$)$^{-1}$. Varying $\alpha_{\rm CO}$ is calculated following \citet{2017MNRAS.470.4750Accurso+}.  Objects with optical spectroscopic coverage smaller than 2 kpc are in gray. The median and $\pm 1\,\sigma$ difference of the two molecular gas mass estimates ($y-x$) for objects with spectroscopic coverage larger than 2 kpc and the whole sample (in parentheses) are shown in the lower-right corner of each panel.}
\label{fig1}
\end{figure*}

\section{Data} \label{sec2}

\subsection{AGN Sample}\label{sec2.1}

This study uses the catalog of broad-line (type~1) AGNs analyzed by \citep{2019ApJS..243...21Liu+}, who performed detailed spectral decomposition of $z<0.35$ galaxies and quasars from the seventh data release \citep[DR7;][]{2009ApJS..182..543Abazajian+} of the Sloan Digital Sky Survey \citep[SDSS;][]{2000AJ....120.1579York+}.  As in \citet{Zhuang&Ho2020}, we study sources having sufficiently high signal-to-noise ratio to permit a clear spectral classification based on narrow emission-line intensity diagnostics, and, in the case of \Ha\ and \Hb, we further require that their signal-to-noise ratio exceeds 5.  We select galaxies classified as hosting Seyfert nuclei and that have a ratio of narrow \Ha\ to \Hb\ larger than 3.1, the canonical intrinsic value for AGNs \citep{2006agna.book.....Osterbrock&Ferland}.  To mitigate against differential aperture effects and possible artificial correlations induced by distance, we focus only on the subset of 453 sources that span the narrow redshift range $0.3<z<0.35$, henceforth dubbed the ``$z=0.3$'' type~1 AGN sample.  The fluxes of the \OIII~$\lambda5007$ line and the narrow component of the \Ha\ and \Hb\ lines are taken directly from the catalog of \citet{2019ApJS..243...21Liu+}, which also provides BH masses estimated from broad \Ha, using the method of \citet{2005ApJ...630..122Greene&Ho}.  The sample consists of powerful AGNs with bolometric luminosities ranging from $L_{\rm bol} \approx 10^{44.3}$ to $10^{47.4}$ erg~s$^{-1}$, which, for a canonical radiative efficiency of 0.1, correspond to BH mass accretion rates of $\dot{M}_{\rm BH} = 0.03-39$~$M_{\sun}$~yr$^{-1}$.  \citet{Zhuang&Ho2020} derived the total stellar masses ($M_*$) of the host galaxies using the empirical scaling relation between BH mass and $M_*$ recently calibrated by \citet{2019arXiv191109678Greene+}. They performed new measurements of \OII~$\lambda3727$, which, in combination with \OIII~$\lambda5007$ and estimates of the gas-phase metallicity, furnish SFRs following the methodology of \citet{2019ApJ...882...89Zhuang&Ho}.

\subsection{Molecular Gas Mass Estimates}\label{sec2.2}

A central part of our analysis requires access to estimates of the gas content---preferably the molecular component---for the AGN host galaxies.  Direct measurements of the molecular gas usually rely on observations of the CO line, which are time-consuming and presently unavailable for the kind of AGN sample needed for the current study.  Estimates of gas masses for AGNs can be obtained indirectly through modeling of the thermal dust emission \citep[e.g.,][]{2018ApJ...854..158Shangguan+, 2019ApJ...873...90Shangguan&Ho}, but it remains challenging to access appropriate far-IR observations to construct the kind of AGN sample necessary for our purposes.

\citet{2019ApJ...884..177Yesuf&Ho} proposed an effective, new empirical method to estimate molecular gas mass from dust extinction. Combining the nebular dust extinction traced by the narrow \Ha/\Hb\ Balmer decrement and gas-phase metallicity, molecular gas masses can be predicted to within $\sim$0.23 dex scatter compared to values derived directly from CO measurements. The scaling relation of \citet{2019ApJ...884..177Yesuf&Ho} depends on the choice of the CO-to-H$_2$ conversion factor ($\alpha_{\rm CO}$).  For a constant Galactic value of $\alpha_{\rm CO}=4.35 \ M_{\sun}$~(K~km~s$^{-1}$~pc$^2$)$^{-1}$,

\begin{equation} \label{eq1}
\begin{split}
& \log (M_{\rm H_2}/M_{\sun}) =  (8.01 \pm 0.11) + \\
& (0.45 \pm 0.10) (A_V / {\rm mag}) + (2.43 \pm 0.42) \log Z, 
\end{split}
\end{equation}

\noindent
while for the varying $\alpha_{\rm CO}$ from \citet{2017MNRAS.470.4750Accurso+}, which primarily depends on gas-phase metallicity with a secondary dependence on the offset from the star-forming galaxy main sequence,

\begin{equation} \label{eq2}
\begin{split}
& \log (M_{\rm H_2}/M_{\sun}) =  (8.27 \pm 0.11) + \\
& (0.38 \pm 0.13) (A_V / {\rm mag}) + (1.44 \pm 0.34) \log Z.
\end{split}
\end{equation}

\noindent 
Here, $A_V$ is the nebular $V$-band dust extinction measured within the SDSS fiber, and $\log Z = 12 + \log {\rm (O/H)} - 8.8$ is the metallicity estimated from the $M_*-Z$ relation of \citet{2004ApJ...613..898Tremonti+}, as parameterized by \citet{2008ApJ...681.1183Kewley&Ellison}. 

The \citet{2019ApJ...884..177Yesuf&Ho} technique was calibrated against a sample of star-forming galaxies.  How reliably can it be applied to the host galaxies of AGNs, particularly type~1 sources?   To address this issue, we compare the molecular gas masses predicted from dust extinction and metallicity with molecular gas masses derived from  CO measurements, using a heterogeneous sample of low-redshift AGNs with available data from the literature.  The literature sources include the Palomar-Green (PG) quasars studied by \citet{2020ApJS..247...15Shangguan+}, the Hamburg/ESO (HE) quasars studied by \citet{2007A&A...470..571Bertram+} and \citet{2017MNRAS.470.1570Husemann+}, the IR-luminous quasars from \citet{2012ApJ...750...92Xia+}, the Seyfert~2 galaxies included as part of the xCOLD GASS survey \citep{2017ApJS..233...22Saintonge+}, and more luminous type 2 quasars from \citet{2012ApJ...753..135Krips+} and \citet{2013MNRAS.434..978Villar-Martin+}.   Cross-matching these samples having CO observations with the SDSS-based type~1 AGN catalog of \citet{2019ApJS..243...21Liu+} and the type~2 AGN catalog from the MPA-JHU database\footnote{http://www.strw.leidenuniv.nl/$\sim$jarle/SDSS/ \\ http://www.mpa-garching.mpg.de/SDSS/DR7/} \citep{2004ApJ...613..898Tremonti+} allows us to locate Balmer decrement measurements and stellar masses\footnote{For type~1 AGNs, stellar masses are estimated from their BH masses using the calibration in \citet{2019arXiv191109678Greene+}. For five type~2 AGNs without stellar mass measurements from the MPA-JHU catalog, we assume $M_*=10^{11}\,M_{\sun}$.} for 22 type~1 and 20 type~2 AGNs.  An additional 10 type~1 AGNs have optical data from the BAT AGN Spectroscopic Survey \citep[][]{2017ApJ...850...74Koss+}. We analyzed these using the quasar spectral fitting code \texttt{PyQSOFIT} \citep{2018ascl.soft09008G} to measure their narrow \Ha\ and \Hb\ fluxes.  In total, we have 32 type~1 and 20 type~2 AGNs. They have redshifts up to 0.34 and cover nearly 5 orders of magnitude in $L_{\rm bol}$ ($10^{41.4} - 10^{46.3}$ erg~s$^{-1}$), overlapping the range of the majority (93\%) of our $z=0.3$ type~1 AGNs. 

As in \citet{2019ApJ...884..177Yesuf&Ho}, we derive molecular gas masses assuming two sets of CO-to-H$_2$ conversion factors. In the case of a variable $\alpha_{\rm CO}$, we use the calibration of \citet{2017MNRAS.470.4750Accurso+}\footnote{We use the IR luminosity to estimate SFR, which is needed to calculating $\alpha_{\rm CO}$. For objects without IR luminosity, the CO(1--0) luminosity is used to predict the IR luminosity using the scaling relation given in \cite{2020ApJS..247...15Shangguan+}.}, whereas for the case of constant $\alpha_{\rm CO}$, we choose one of two fixed values, depending on the IR luminosity ($L_{\rm IR}$) of the object. For $L_{\rm IR}\ll10^{12}\,L_{\sun}$, which applies to the PG quasars, HE quasars, and Seyfert 2 galaxies, we adopt $\alpha_{\rm CO} = 3.1\ M_{\sun}$~(K~km~s$^{-1}$~pc$^2$)$^{-1}$, a value found to be appropriate for low-redshift quasars \citep{2020ApJS..247...15Shangguan+}; for IR-luminous quasars and type~2 quasars, characterized by $L_{\rm IR}\ga10^{12}\,L_{\sun}$, we select $\alpha_{\rm CO} = 0.8\ M_{\sun}$~(K~km~s$^{-1}$~pc$^2$)$^{-1}$, a value recommended for ultraluminous IR galaxies \citep{1998ApJ...507..615Downes&Solomon}. 

We use the Kaplan-Meier estimator, as implemented in the {\tt Python} package \texttt{lifelines}\footnote{https://lifelines.readthedocs.io/en/latest/}, to calculate the median and $16\%-84\%$ interval of the molecular gas mass, including the CO non-detections.  The extinction-based molecular gas masses, either for the varying or constant $\alpha_{\rm CO}$, generally provide reasonable estimates of the true gas masses based on actual CO measurements, albeit with large scatter (Figure~\ref{fig1}). One of sources of the scatter presumably arises from the mismatch in spatial coverage between the CO measurements and the optical spectroscopic data, which generally cover just the central region of the host galaxy.  A larger value of dust extinction is often found in the central region of a galaxy \citep[e.g.,][]{2019ApJ...887..204Jafariyazani+}. After excluding 15 objects whose optical spectra were acquired with a slit width smaller than 2 kpc, the scatter for the case of constant $\alpha_{\rm CO}$ is significantly reduced from 0.83 to 0.50 dex.  The median differences between the two molecular gas mass estimates are much smaller than the $\pm1\,\sigma$ scatter.  In spite of the small-number statistics, we are encouraged by the results of this comparison.  For concreteness, all subsequent analysis will use Equation~\ref{eq1} to predict molecular gas mass.  We note that we have probably overestimated the real uncertainty of converting nebular extinction and metallicity to molecular gas mass.  The heterogeneous origin of the CO measurements from different telescopes and observations likely leads to calibration systematics and contributes to the observed scatter of $\sim$0.5 dex.

\begin{figure}[t]
\centering
\includegraphics[width=0.5\textwidth]{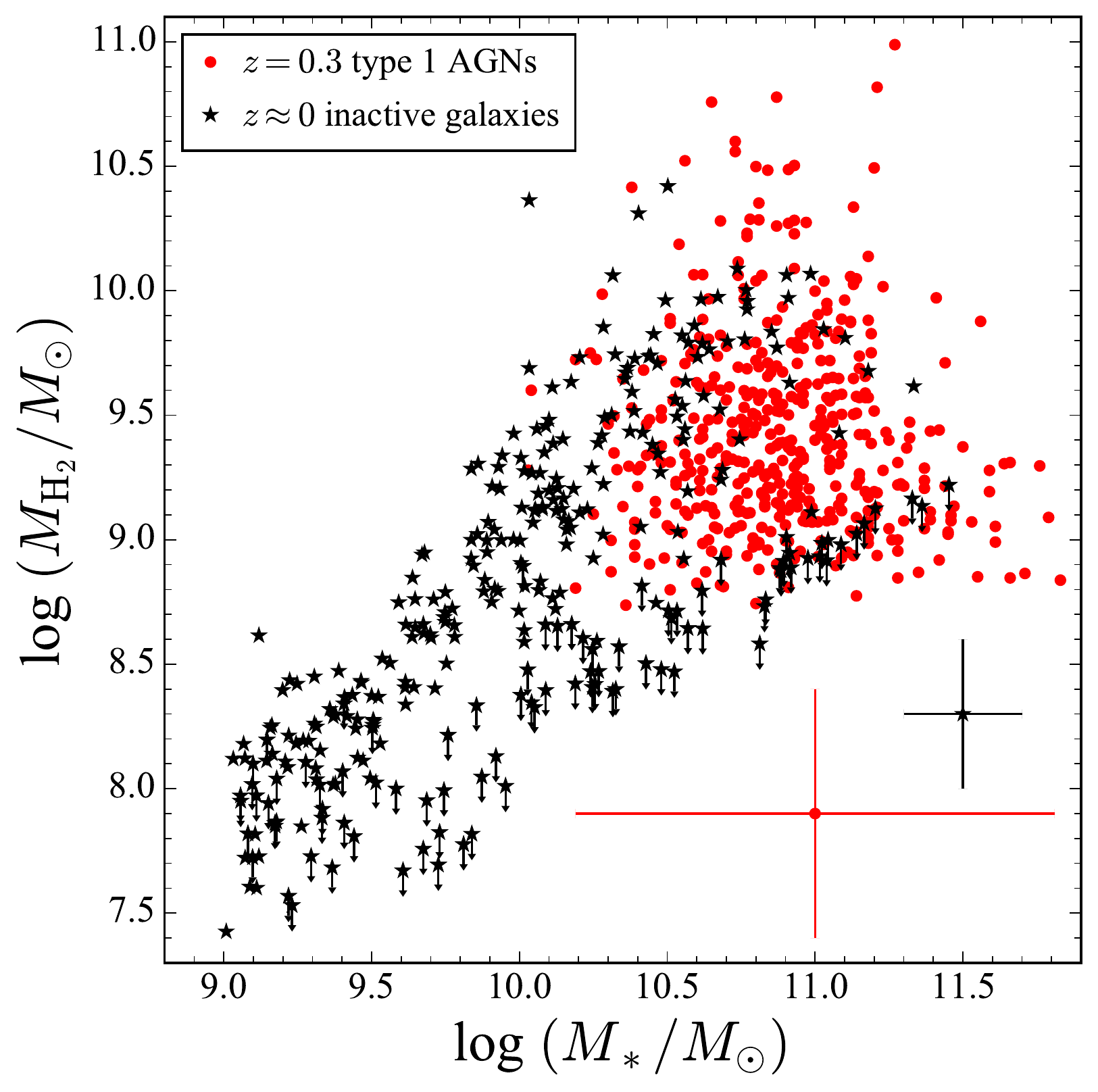}
\caption{Molecular gas mass versus stellar mass for our sample of $z=0.3$ type~1 AGNs (red circles) and $z\approx0$ inactive galaxies (black stars) from the xCOLD GASS sample \citep{2017ApJS..233...22Saintonge+}. Typical uncertainties are given in the lower-right corner.}
\label{fig2}
\end{figure}

\begin{figure*}[t]
\centering
\includegraphics[height=0.5\textwidth]{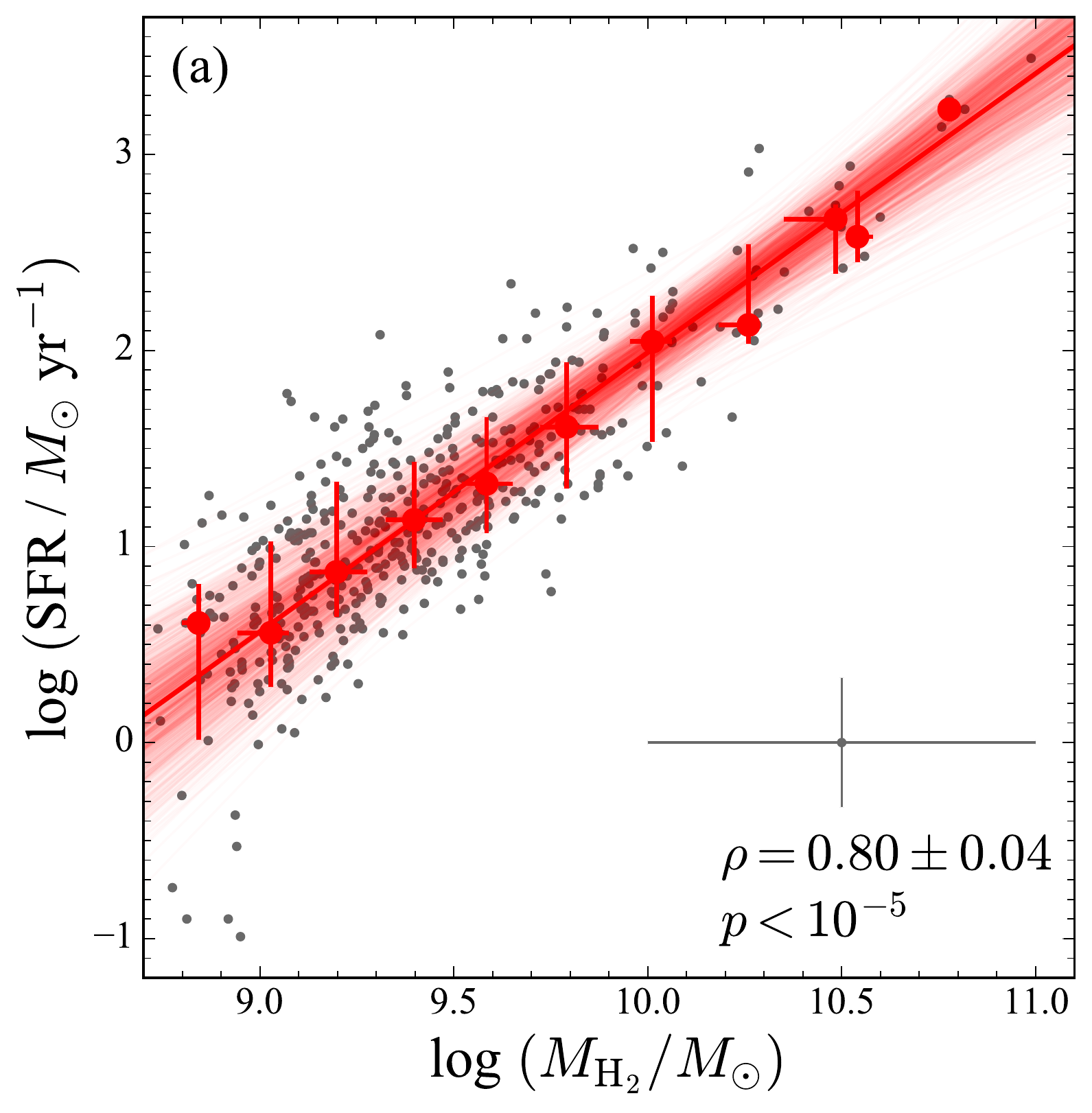}
\includegraphics[height=0.5\textwidth]{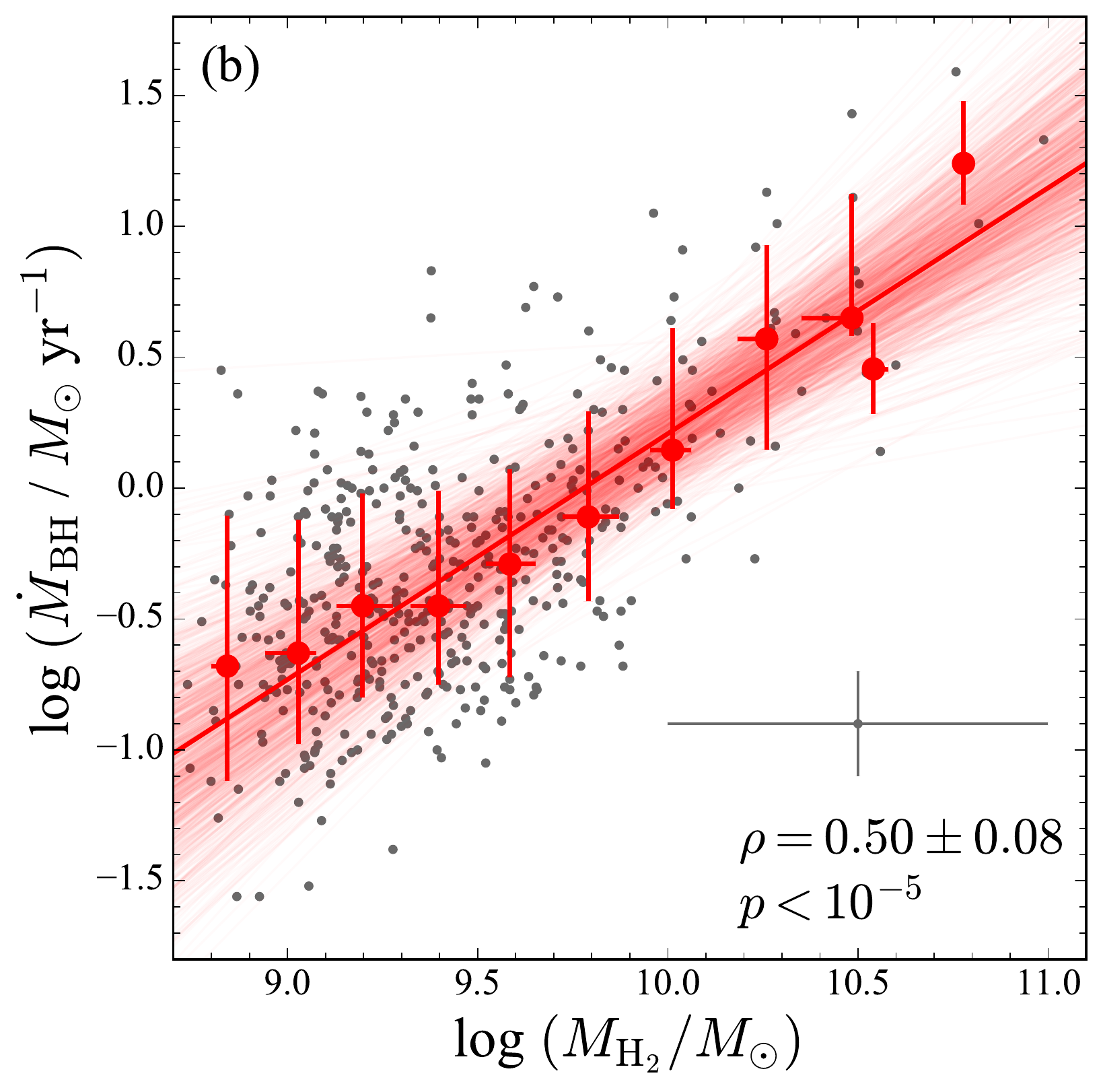}
\caption{(a) Star formation rate and (b) BH accretion rate versus molecular gas mass for our $z=0.3$ type~1 AGN sample. Small gray points represent individual objects,  with typical errorbars indicating $1\,\sigma$ uncertainty shown at the lower-right corner. Large red points indicate the median value in bins of 0.2 dex in $M_{\rm H_{2}}$, with errorbars indicating 16th and 84th percentile. Fitting to the medians are visualized using the red lines.  The median Spearman correlation coefficient, its standard deviation, and $p$-value for bootstrap resampling 500 times using samples of 100 objects are given in the lower-right corner of each panel.}
\label{fig3}
\end{figure*}

\begin{figure}[t]
\centering
\includegraphics[height=0.5\textwidth]{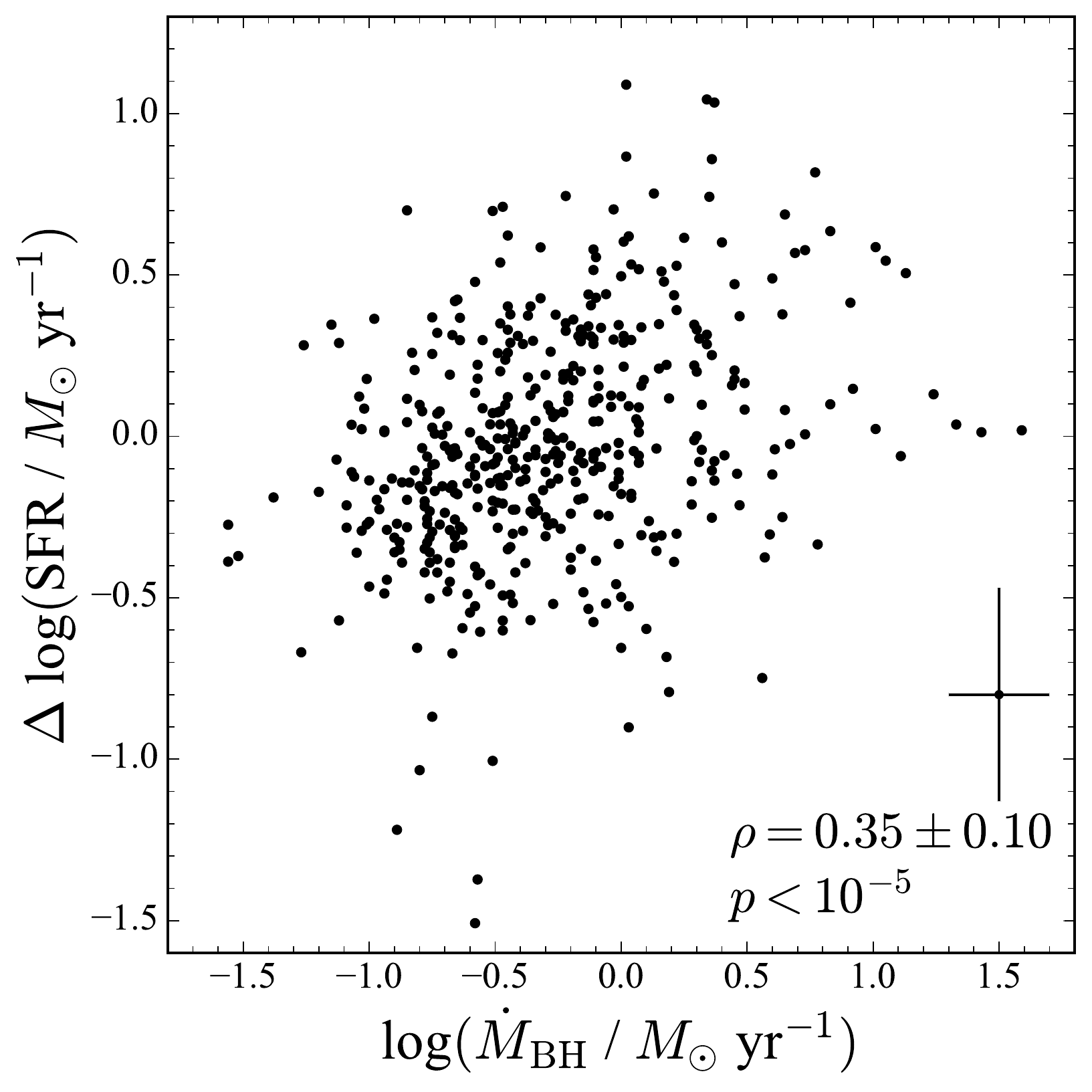}
\caption{Residual SFR versus BH accretion rate after removing the dependence of SFR on $M_{\rm H_2}$ using Equation~\ref{eq3}, for the sample of $z=0.3$ type~1 AGNs.  The residual SFR is defined as $\Delta{\rm SFR} = {\rm SFR} -(1.44  \log{M}_{\rm H_2}-12.37)$. We quantify the uncertainty of subtracting the correlation between SFR and $M_{\rm H_2}$ using 500 fitting trials randomly selected from the posterior distribution of the fit (light red lines in Figure~\ref{fig3}a). The median Spearman correlation coefficient, its standard deviation, and $p$-value are shown in the lower-right corner.}
\label{fig4}
\end{figure}

\section{Results}\label{sec3}

\subsection{Active and Inactive Galaxies Have Similar Gas Content}\label{sec3.1}

The molecular gas content of a galaxy varies with its stellar mass \citep{2016MNRAS.462.1749Saintonge+}. Figure~\ref{fig2} shows that the molecular gas masses for our $z=0.3$ type~1 AGNs are generally consistent with those for nearby inactive, star-forming galaxies from the xCOLD GASS sample \citep{2017ApJS..233...22Saintonge+}. This implies that $z=0.3$ type~1 AGNs have similar molecular gas content as nearby normal galaxies, echoing previous cold gas surveys of low-$z$ quasars \citep[e.g.,][]{2007A&A...470..571Bertram+, 2013MNRAS.434..978Villar-Martin+, 2020MNRAS.498.1560Jarvis+, Shangguan+2020}. Using the same technique as we, \citet{Yesuf&Ho2020} reached a similar conclusion regarding nearby Seyfert 2 galaxies.  The existence of a large gas reservoir in AGNs supports the idea that, instead of instantaneously removing the interstellar medium and suppressing star formation in their host galaxies, AGN feedback may operate over a longer timescale \citep[e.g.,][]{2014MNRAS.444.2355Costa+, 2017NatAs...1E.165Harrison, 2018NatAs...2..198Harrison+}. 

Figure~\ref{fig2} might give the impression that our AGN sample contains no gas-poor objects whatsoever.  There are, although not many. The selection criteria of \citet{Zhuang&Ho2020} required that narrow \Ha/\Hb\ $> 3.1$, to exclude objects with unphysical Balmer decrements that can arise from poor spectral decomposition. This requirement, although very stringent, only excluded $\sim 3\%$ (13/466) of the parent sample of $z = 0.3$ type 1 AGNs.  Moreover, Equation~\ref{eq1} can only probe $M_{\rm H_2} \gtrsim 10^{8.6}\,M_{\sun}$ for objects with $M_*>10^{10.5}\,M_{\sun}$, even when the dust extinction is zero.  Thus, while we miss gas-poor systems, we do not miss many, and they do not affect our main conclusions.  The overall consistency between our results and independently derived gas masses lends confidence that the extinction-based method can be applied to estimate molecular gas masses in type~1 AGNs.

\subsection{SFR and $\dot{M}_{\rm BH}$ Are Intrinsically Correlated}\label{sec3.2}

\begin{figure*}[t]
\centering
\includegraphics[width=0.7\textwidth]{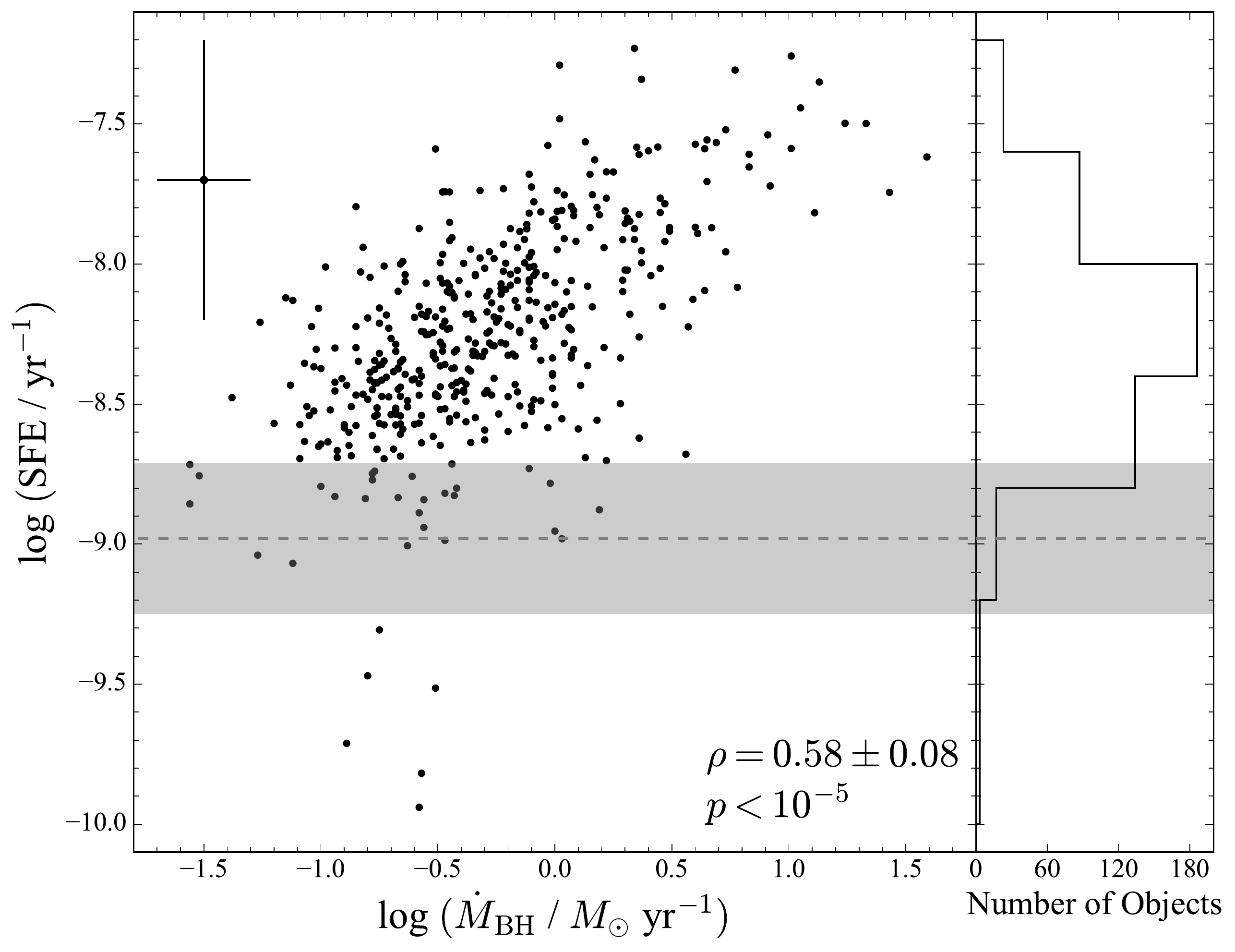}
\caption{Star formation efficiency (SFE) versus BH accretion rate for $z=0.3$ type~1 AGNs, with the histogram of SFE given in the right panel. Typical uncertainties are shown in the upper-left corner, and the median Spearman correlation coefficient, its standard deviation, and $p$-value derived from bootstrap resampling 500 times using samples of 100 objects are shown in the lower-right corner. Gray dashed horizontal line represents the mean SFE of nearby main-sequence galaxies from \citet{2017ApJS..233...22Saintonge+}, with the shaded area indicating the $\pm1\,\sigma$ range.}
\label{fig5}
\end{figure*}

SFR correlates with molecular gas mass \citep[e.g.,][]{2008AJ....136.2846Bigiel+, 2010MNRAS.407.2091Genzel+}, and so, too, does AGN luminosity \citep{2012ApJ...750...92Xia+, 2016ApJ...827...81Izumi+, 2017MNRAS.470.1570Husemann+, Shangguan+2020}.  We also find a very strong positive correlation (Spearman correlation strength $\rho_1=0.80\pm0.04$, $p<10^{-5}$; Figure~\ref{fig3}a)\footnote{Throughout this paper we consider a correlation ``very strong'' when the Spearman's coefficient $\rho \geq 0.8$, ``strong'' when $0.6 \leq \rho < 0.8$, ``moderate'' when $0.4 \leq \rho < 0.6$, ``weak'' when  $0.2 \leq \rho < 0.4$, and ``very weak'' when $\rho < 0.2$.} between SFR and $M_{\rm H_2}$ and a moderately strong positive correlation ($\rho_2=0.50\pm0.08$, $p<10^{-5}$; Figure~\ref{fig3}b) between $\dot{M}_{\rm BH}$ and $M_{\rm H_2}$ for our $z=0.3$ type~1 AGN sample.  To mitigate against the potential effect of the large uncertainties in the molecular gas mass estimates on our correlation analysis, we use bootstrap resampling\footnote{We resample 500 times using samples of 100 objects.  The statistics are stable (almost identical $\rho$ and $p$) for sample sizes larger than 70.} to quantify the uncertainties when performing correlation analysis involving $M_{\rm H_2}$ (i.e. Figures \ref{fig3} and \ref{fig5}). Fits to the medians of the data in Figure~\ref{fig3} using the linear regression code \texttt{linmix} \citep{2007ApJ...665.1489Kelly} give

\begin{equation} \label{eq3}
\begin{split}
{\rm \log (SFR}/M_{\sun}\ {\rm yr}^{-1}) = & (-12.37 \pm 1.66) + \\
& (1.44 \pm 0.16) \log({M}_{\rm H_2}/M_{\sun})
\end{split}
\end{equation}

\noindent and 

\begin{equation} \label{eq4}
\begin{split}
 \log (\dot{M}_{\rm BH}/M_{\sun}\ {\rm yr}^{-1}) = & (-9.47 \pm 2.13) + \\
& (0.97 \pm 0.21) \log({M}_{\rm H_2}/M_{\sun}). 
\end{split}
\end{equation}

Two important inferences can be drawn from these results.  On the one hand, the relation between SFR and molecular gas mass observed in our sample of type~1 AGNs implies that even in powerful AGNs the interstellar medium of their host galaxies behaves essentially normally, insofar as their ability to form stars is concerned.  On the other hand, the existence of an empirical relation between BH accretion rate and total molecular gas mass suggests that there is a physical link between the gas supply of the host galaxy on global scales and the fuel reservoir for the AGN on circumnuclear scales.  This echoes the results from the studies of \cite{2017MNRAS.470.1570Husemann+} and \cite{Shangguan+2020}, but places them on much firmer statistical footing because of the unprecedented size and homogeneity of our sample.

We return to the main issue that triggered this study, one that has motivated many similar studies in the literature (Section~\ref{sec1}).  To what extent does the SFR of the host galaxy truly correlate with the BH accretion rate (or AGN luminosity)?  Taken at face value, the full parent sample of all $z \leq 0.35$ type~1 AGNs certainly exhibits a dramatic correlation between SFR and $\dot{M}_{\rm BH}$ \citep[their Figure~5a]{Zhuang&Ho2020}.  However, this can be misleading, for an artificial correlation can be induced by redshift and stellar mass.  After mitigating these effects by limiting the analysis to the subsample isolated to $z=0.3$, \citet[their Figure~5b]{Zhuang&Ho2020} show that a highly significant positive correlation ($\rho_3=0.68$, $p<10^{-5}$) persists, with no evident dependence on stellar mass.  Our present study reveals yet another factor of concern, one that hitherto has been underappreciated.  In light of the separate ${\rm SFR}-M_{\rm H_2}$ and $\dot{M}_{\rm BH}-M_{\rm H_2}$ correlations discussed above (Figure~\ref{fig3}), seeking any additional, intrinsic link between SFR and $\dot{M}_{\rm BH}$ first should remove the mutual dependence on $M_{\rm H_2}$.  This is obviously a difficult and demanding task, given the myriad requirements that the sample must satisfy.  It also highlights the dangers of over-interpreting any casual presentation of the SFR-$\dot{M}_{\rm BH}$ relation without considering this factor. Although our estimates of molecular gas mass, based on the Balmer decrement and metallicity from the stellar mass-metallicity relation, are admittedly crude \citep{2019ApJ...884..177Yesuf&Ho} and necessarily indirect, they afford us the opportunity to investigate the partial correlation between SFR and $\dot{M}_{\rm BH}$ after removing the dependence of SFR on $M_{\rm H_2}$.  Since SFR correlates more strongly with $M_{\rm H_2}$ than it does with $\dot{M}_{\rm BH}$, while $\dot{M}_{\rm BH}$ does not ($\rho_1>\rho_3>\rho_2$), we only consider the result of removing the dependence of SFR on $M_{\rm H_2}$. Toward this end, to remove the dependence of SFR on $M_{\rm H_2}$, we subtract from each observed SFR the value predicted from $M_{\rm H_2}$ using Equation~\ref{eq3}, to obtain the residual SFR: $\Delta {\rm SFR} = {\rm SFR} -(1.44  \log{M}_{\rm H_2} - 12.37)$.  We estimate the uncertainties of this subtraction using the 500 random realizations from the posterior distribution of the fit (thin red lines in Figure~\ref{fig3}a).  Figure~\ref{fig4} displays $\Delta$SFR versus BH accretion rate. Intriguingly, even after accounting for the dependence of SFR on $M_{\rm H_2}$,  the residual SFR {\it still}\ significantly correlates with $\dot{M}_{\rm BH}$ ($p<10^{-5}$), albeit more weakly ($\rho=0.35 \pm 0.10$)\footnote{A larger Spearman coefficient is achieved ($\rho=0.53\pm0.04$, $p<10^{-5}$) if the dependence of $\dot{M}_{\rm BH}$ on $M_{\rm H_2}$ is further included.}.  The existence of an intrinsic correlation between $\dot{M}_{\rm BH}$ and SFR suggests that BH accretion in the innermost region somehow communicates with star formation on galactic scales.

\section{Discussion}\label{sec4}
\subsection{Implications for AGN Feedback}\label{sec4.1}
 
Our partial correlation analysis suggests that, after accounting for the common dependence on $M_{\rm H_2}$, an intrinsic relation exists between SFR and $\dot{M}_{\rm BH}$. Far from curtailing star formation, BH accretion evidently is connected somehow with the positive enhancement of star formation activity in the host galaxy.  While we currently cannot say where in the host galaxy the stars form, \citet{Zhuang&Ho2020} present tentative evidence that the star formation occurs predominantly on relatively small (central $\sim 1$ kpc) scales.  

A number of authors have proposed that AGN feedback can exert a positive instead of a negative influence on star formation \citep[e.g.,][]{2005ApJ...635L.121King, 2012MNRAS.427.2998Ishibashi&Fabian, 2013MNRAS.431.2350Ishibashi+}. Positive feedback can be achieved by compression of cold molecular gas by fast outflows generated by AGNs \citep{2013ApJ...772..112Silk, 2015A&A...582A..63Cresci+} and by the direct formation of stars via cooling and fragmentation of the gas inside the outflows themselves \citep{2017Natur.544..202Maiolino+, 2019MNRAS.485.3409Gallagher+}.  While we presently lack the spatial information to make more definitive statements about the nature of the connection between the AGN and star formation, Figure~\ref{fig5} shows that the star formation efficiency (${\rm SFE} \equiv {\rm SFR}/M_{\rm H_2}$) strongly correlates ($\rho=0.58\pm0.08$, $p<10^{-5}$) with the BH accretion rate, with the majority of the AGNs in our sample having SFEs higher than the typical values of galaxies on the star-forming main-sequence in the local Universe \citep{2017ApJS..233...22Saintonge+}. Stars not only form more efficiently in AGN hosts, but their formation efficiency increases with larger BH accretion rate.  

Outflows occur pervasively in AGNs, and the fraction of AGNs with outflow signatures increases with higher \OIII\ luminosity \citep{2018ApJ...865....5Rakshit&Woo}.  Using integral field spectroscopy of 2800 local galaxies, \citet{2019MNRAS.485.3409Gallagher+} find that the SFR inside the outflows positively correlates with the mass outflow rate of the ionized gas. Equation~9 of \citet{2019MNRAS.485.3409Gallagher+} implies that the fraction of the SFR inside outflows increases from $\sim10\%$ at $\dot{M}_{\rm BH}=0.1 \, M_{\sun}$~yr$^{-1}$ to $\sim 80\%$ at $\dot{M}_{\rm BH}=3\, M_{\sun}$~yr$^{-1}$. Galactic outflows contain a large reservoir of molecular gas \citep[e.g.,][]{2010A&A...518L.155Feruglio+, 2019MNRAS.483.4586Fluetsch+, 2020arXiv200613232Fluetsch+}, which occupy a higher proportion in a dense phase compared to the galactic disk \citep{2012A&A...537A..44Aalto+, 2017ApJ...835..265Walter+}.  The  relative fraction of molecular gas to ionized gas in outflows scales with AGN strength \citep{2019MNRAS.483.4586Fluetsch+}. We surmise that the overall rise of SFE with BH accretion rate may be connected with positive feedback from AGN-driven outflows.

We close with a note of clarification. \citet{Shangguan+2020} recently studied the physical properties of the host galaxies of 40 PG quasars for which they could measure their molecular gas content through CO observations and SFRs from decomposition of the full ($1-500\,\mu$m) IR spectral energy distribution. The authors find that SFR correlates strongly with AGN luminosity, but that the correlation disappears after taking into account the mutual dependence on molecular gas mass (their Figure 8). We attribute this apparent discrepancy with the results of this paper to small-number statistics.  We can qualitatively reproduce the results of \citep{Shangguan+2020} by performing bootstrap resampling 500 times of subsets of 40 objects  (to match the sample size of \citealt{Shangguan+2020}) selected from our parent sample of 453 sources.  There is a 15\% probability that the randomly drawn samples achieve a Spearman $p$-value $>0.01$.  

\subsection{Caveats}\label{sec4.2}

This study extends to AGNs the formalism of \cite{2019ApJ...884..177Yesuf&Ho}, originally devised for star-forming galaxies, to estimate molecular gas mass using nebular extinction and metallicity. To be sure, the indirect molecular gas masses have significant uncertainties ($\sim$0.5 dex; Figure~\ref{fig1}a), and they should be used with extreme caution for any single individual object.  However, the absence of systematic bias between the molecular gas masses based on CO and those estimated from dust extinction lends confidence that we can apply our method to study the molecular gas content of a large AGN sample. Although our main sample contains a small percentage ($\sim$8\%) of objects with $L_{\rm bol}$ and $M_{\rm H_2}$ larger than those of the calibration sample, we verified that excluding this small subset of objects does not affect our results. 

We suggest that positive AGN feedback might underlie the positive correlation between SFE and $\dot{M}_{\rm BH}$ (Section~\ref{sec4.1}).  Perhaps some alternative process can enhance the host galaxy SFE while simultaneously coupling to the BH accretion rate.  For example, gas-rich galaxy-galaxy interactions can drive gas from the outskirts of the host galaxy to the center, induce intense central star formation, and feed the BH \citep[e.g.,][]{2005ApJ...630..705Hopkins+}.  Unfortunately, the quality of the available SDSS optical images precludes us from obtaining reliable information on the morphology of the sample. The objects are located at a relatively large distance ($z\approx0.3$), compounded by the fact that the prominent type~1 nucleus presents a major source of contamination.  Nevertheless, there is no strong evidence that major mergers play a significant role in triggering BH accretion in quasars (\citealp{2003MNRAS.340.1095Dunlop+, 2011ApJ...726...57Cisternas+, 2016ApJ...830..156Mechtley+, 2019ApJ...877...52Zhao+}; but see counter evidence in \citealp{2020arXiv201000022Marian+}). Studies of the smaller sample of nearby ($z < 0.5$) PG quasars reinforce our conclusions: \cite{Shangguan+2020} and \cite{Xie+2020} report that a sizable fraction of these quasars also exhibit high SFEs, and yet many lack signatures of ongoing or recent major mergers.  Although other possibilities such as minor mergers \citep{2014MNRAS.443..755Husemann+} cannot be ruled out easily, we suspect that major mergers are not responsible for the observed correlation between SFE and $\dot{M}_{\rm BH}$.

Lastly, we note that our AGN sample, intentionally designed to mitigate complications from redshift effects and SDSS fiber coverage, limits our study only to relatively luminous AGNs at $z\approx 0.3$. It is possible that BH accretion has less of an impact on the star formation properties of the host galaxies of less luminous AGNs, which exhibit a lower incidence of outflows (e.g., \citealt{2018ApJ...865....5Rakshit&Woo}).  While elevated SFEs can be found in quasars at $z \lesssim 0.3$ \citep{2017MNRAS.470.1570Husemann+, Shangguan+2020}, they are not as prevalent as in the sample considered here.

\section{Summary}\label{sec5}

Using a large sample of 453 $0.3<z<0.35$ type~1 AGNs from \citet{Zhuang&Ho2020}, we apply the molecular gas mass estimator from \citet{2019ApJ...884..177Yesuf&Ho} to study the link between BH accretion and SFR after accounting for the dependence of these quantities on redshift and molecular gas mass. We summarize this study as follows:

\begin{enumerate}
\item{We collect 32 type~1 and 20 type~2 AGNs with archival CO observations and Balmer decrement from optical spectroscopy and show that the combination of dust extinction and metallicity prescribed by \citet{2019ApJ...884..177Yesuf&Ho} provides fairly accurate molecular gas mass estimates for AGNs.}

\item{Applying the molecular gas estimator to our $z=0.3$ type~1 AGNs, we find that these objects have similar gas content as nearby inactive galaxies, which suggests that AGNs do not remove the cold gas content of their host galaxies instantaneously.}

\item{We find that both SFR and BH accretion rate correlate with molecular gas mass. The observed strong correlation between SFR and $\dot{M}_{\rm BH}$ is exaggerated by their mutual dependence on $M_{\rm H_2}$.}

\item{After removing the dependence of SFR on $M_{\rm H_2}$, SFR and $\dot{M}_{\rm BH}$ are still weakly correlated, which suggests that BH accretion in the innermost region is linked to star formation on galactic scales.}

\item{We find a strong correlation between star formation efficiency and BH accretion rate, which can be interpreted as evidence for positive AGN feedback.}
\end{enumerate}

\vspace{5mm}
\software{Astropy \citep{2013A&A...558A..33A, 2018AJ....156..123A},  Matplotlib \citep{Hunter:2007}, Numpy \citep{numpy}, PyQSOFIT \citep{2018ascl.soft09008G} Scipy \citep{scipy} }

\acknowledgments

This work was supported by the National Science Foundation of China (11721303, 11991052) and the National Key R\&D Program of China (2016YFA0400702).  We thank the referee for constructive comments and suggestions that improved the paper. We are grateful to Shu Wang and Hengxiao Guo for help on running PyQSOFIT and Ruancun Li for double-checking our spectral fitting. M.-Y. Zhuang thanks Gregory Herczeg for holding a writing class, and Bitao Wang and Yu-Ming Fu for help with editing.

\end{document}